\documentclass[epsf]{elsart}
\usepackage{epsfig}

\usepackage{graphicx}
\usepackage{longtable}
\usepackage{dcolumn}
\usepackage{bm}

\def\nimol{[$\mbox{As}@\mbox{Ni}_{12}@\mbox{As}_{20}]^{-3}$ }
\def\nimolo{$\mbox{As}@\mbox{Ni}_{12}@\mbox{As}_{20}$ }
\def\nicage{As@Ni$_{12}$ }
\def\ascage{As$_{20}$ }
\def\as4{As$_{4}$ }
\def\cmin {cm$^{-1}$ }

\begin{document}

\begin{frontmatter}

\title{Stability of As$_n$ [n=4, 8, 20, 28, 32, 36, 60] Cage Structures}

\author{Tunna Baruah$^{1,2}$, Mark R. Pederson$^2$, Rajendra R. Zope$^{3}$, and M. R. Beltr\'an$^4$}

\address{$^1$ Department of Physics, Georgetown University, Washington DC, 20057}

\address{$^2$Center for Computational Materials Science, Naval Research 
Laboratory, Washington DC 20375-5345}

\address{$^3$School of Computational Sciences, George Mason University, Fairfax,
VA 22030}

\address{$^4$Instituto de Investigaciones en Matriales, Universidad Nacional Aut\'onoma de M\'exico, M\'exico D.F. A.P. 70-360, C.P. 04510} 

\date{\today}

\begin{abstract}

We  present all-electron density functional study of the geometry, electronic structure,
vibrational modes, polarizabilities as well as the infrared and Raman spectra
of fullerene-like arsenic cages.
The stability of As$_n$ cages for sizes 4, 8, 20, 28, 32, 36, and 60 wherein
each As atom is three-fold coordinated is examined.
We find that all  the cages studied are vibrationally stable and
while all the clusters are energetically stable with respect to isolated
arsenic atoms,
only As$_{20}$ is energetically stable against dissociation into As$_4$.
We suggest that
the Raman spectra might be a means for observing the As$_{20}$ molecule in gas phase.

\end{abstract}

\end{frontmatter}

 Highly symmetric molecular cages are of great interest due to their inherent symmetry and 
bonding\cite{Muller}. The highly studied carbon fullerenes \cite{smalley,KLFH,Fowler_book}
are prime examples of this class of materials. Apart
from their symmetry they also show unusual properties such as superconductivity in the solid
phase \cite{Haddon92}
and storage capacity which can have potential applications in nanotechnology.  For more than a decade
considerable effort has been paid to the possibility of creating fullerene-like structures with
elements other than carbon.
Some studies have shown Si and Ge clusters in the gas phase
\cite{sicr,KRJ,MGST,SC,CJSH} form cages but their structures are still controversial.
Recently Moses and coworker \cite{MFE} have successfully synthesized a highly symmetric
onion-like cage formed by an icosahedral \nicage cluster with an As atom at the center. The resulting
13-atom cluster is further encapsulated by a dodecahedral fullerene-like \ascage.
This report opens up the possibility of formation of other possible cage-like As$_n$ clusters
in the laboratory.  In the present study, we examine the possibility for obtaining stable
As cages for sizes upto sixty atoms.

 The clusters of other isoelectronic elements nitrogen and phosphorus are well studied. Nitrogen 
clusters are studied primarily because of their high energy density. It may be pointed out that 
although small N clusters are stable, the higher size clusters tend to break up  into smaller 
clusters.  Owens \cite{O} has studied N$_4$, N$_8$,N$_{10}$, N$_{12}$ and N$_{20}$ clusters 
theoretically using density functional theory (DFT). He has shown that all the nitrogen clusters have 
high-energy density and will decompose into N$_2$ releasing a large amount of energy with
highest release for N$_8$.  A number of phosphorus clusters with closed-shell structure are 
proposed to be more energetic than P$_4$ \cite{phosphorus,JGHP,HT,BBKO} theoretically.  While the 
N$_{20}$ clusters are highly unstable, the P$_{20}$ clusters are found to have higher stability. 
The prediction for P$_{20}$ and the experimental existence of the \nimolo are indicative of the 
possible existence of large As clusters.

 The stability of the As$_n$ clusters were theoretically studied by Shen and Schaefer for sizes 2,
4, 12, and 20 \cite{As20}. However, they concluded that As$_{20}$ would be energetically competitive 
with five As$_4$ clusters. Our recent calculations done on \nimolo and As$_{20}$ indicates that 
As$_{20}$ is stable both electronically and vibrationally.  Its most favored dissociation channel 
is the As tetramer as predicted by Shen and Schaefer \cite{As20}. In this work we report our 
study of other possible As cages and their electronic and vibrational stability. It is possible 
that the ground state structure of the As clusters considered here may not conform to a cage-like 
structure.  However, we have restricted our study only to cage structures and also to the ones 
which may break up into integral number of tetramers. The focus is on finding vibrationally stable
symmetric cages and the energy of dissociation into tetramers. Also we have restricted the cage 
structures in which the As atoms are in  3-fold coordination as seen in the experimentally 
obtained As shell of the \nimol cluster. In this case,
the As atoms are sp$^3$ hybridized with a lone pair and 3 half-filled orbitals which
form $\sigma$ bonds with its nearest  neighbors. Shen and 
Schaefer have shown that the tetrahedral As$_4$ is a very stable cluster.
This follows directly from the considerations that As atoms are sp$^3$ hybridized and prefer 
3-fold coordination with lone pairs. 
A similar feature was also seen in \nimolo cluster \cite{BZRP}.
Moreover, the twenty valence electrons associated with 
the As$_4$ molecule is known to correspond to a magic number in metallic clusters and this
aspect may further contribute to the stability of the tetramer \cite{knight}.

Our density functional theory \cite{DFT1,DFT2} based  calculations were performed at the all-electron 
level within the generalized gradient 
approximation (GGA) \cite{PBE}  to describe the exchange-correlation effects. The calculations 
have been performed using the NRLMOL package \cite{NRLMOL1,NRLMOL2,NRLMOL3} which employs a 
Gaussian basis set 
where the exponentials are optimized for each atom \cite{PP1}.  The basis for the As contains 
7s, 6p, and 
4d type contracted Gaussians along with a d-type polarization function. The  package also 
employs a variational mesh to calculate the integrals accurately and also efficiently. 
The Hartree potential is calculated analytically. 
The self-consistency cycle was carried out till the energies converged to 1.0 
$\times$ 10$^{-6}$ Hartree. The symmetry restricted geometry optimization was carried out
using the LBFGS scheme 
till the forces were smaller than 0.001 a.u.. 
The vibrational frequencies are calculated by introducing 
small perturbation to the equilibrium geometry in the Cartesian directions for all atoms 
and calculating forces. From these, the dynamical matrix is calculated by finite difference 
method, diagonalization of which  yields the frequencies \cite{PP}.

\begin{figure}[t]
\epsfxsize=3.5in
\centerline{\epsfbox{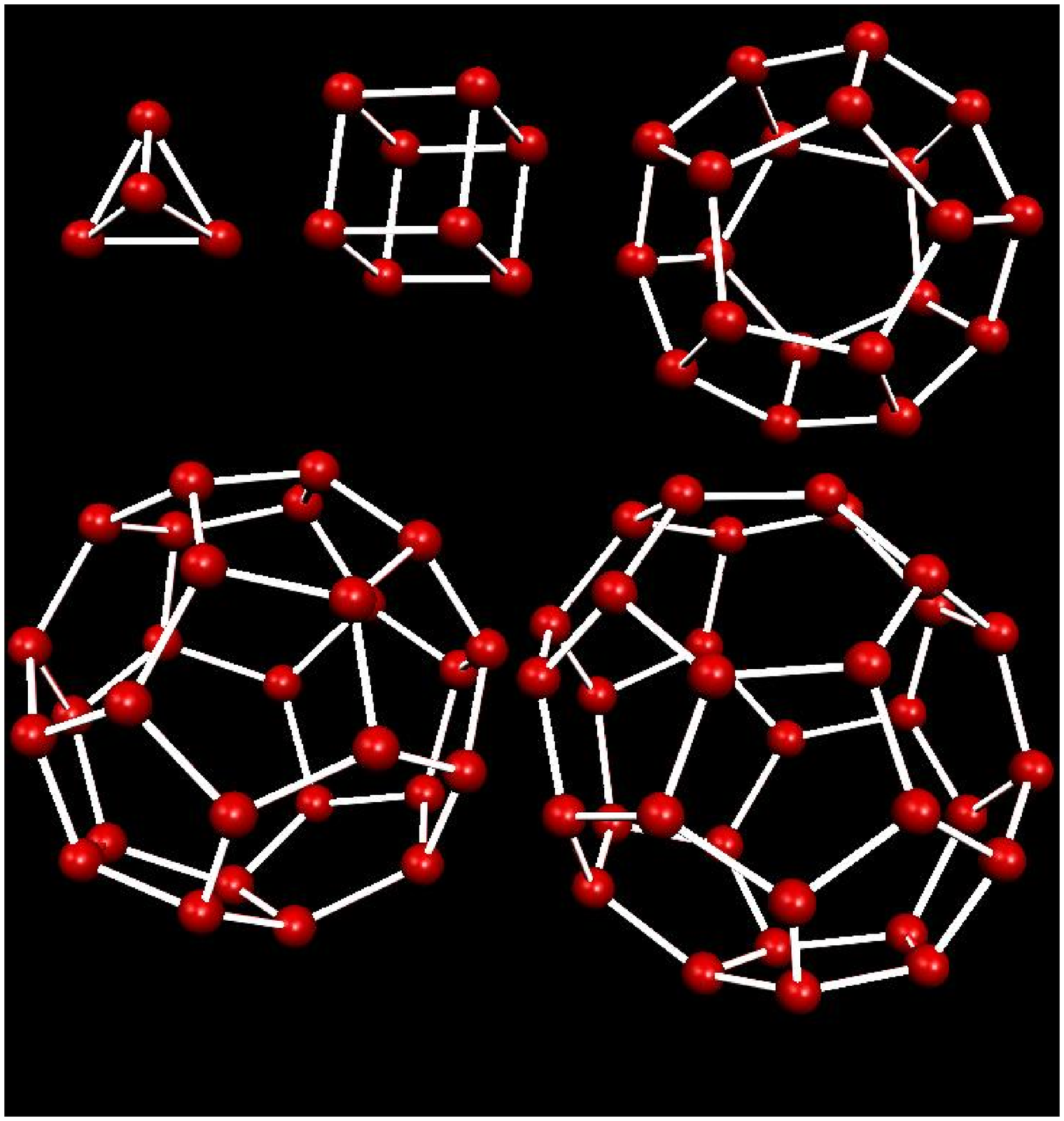}}
\caption{\label{fig1}  Structures of As$_{4}$,  $As_{8}$, As$_{20}$,  As$_{28}$, and As$_{32}$ 
}
\end{figure}

\begin{figure}[b]
\epsfxsize=4.5in
\epsfysize=3.5in
\centerline{\epsfbox{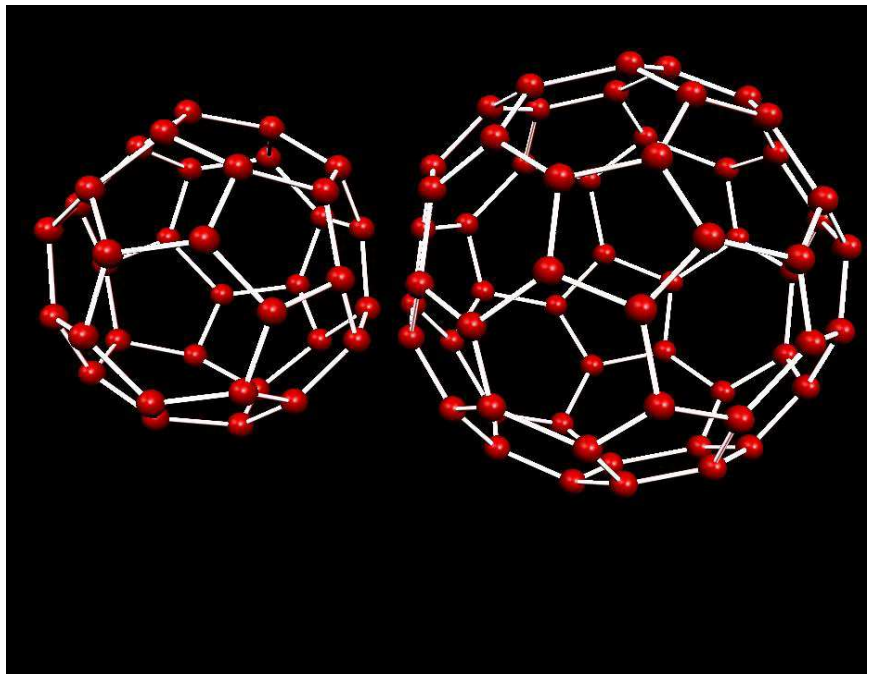}}
\caption{\label{fig2}  Structures of As$_{36}$  and As$_{60}$.}
\end{figure}

We present the  optimized structures 
of the As$_{4,8,20,28,32,36,60}$  fullerene-like cages in Figs. \ref{fig1} and \ref{fig2}.
The range of nearest neighbor As-As bond lengths, the bond angles, and the symmetry of the molecules 
are listed in Table \ref{table:str}. The bond lengths of the cages lie between 2.43 to 2.74 \AA.
The As-As bond length in the experimentally observed \nimolo cluster is 2.75 \AA which is close
to the largest bond-length seen in the As cages. The bond angles for the smallest cluster is 60$^o$
which increases to the range 104-130$^o$ for the larger cages. The As$_4$ and As$_8$ clusters
have T$_d$ and O$_h$ point symmetry. The bond lengths and angles do not vary in these structures due to
their symmetry. The As$_{20}$ has a  dodecahedron structure made up of 12 pentagons. The 28-atom
cluster also has T$_d$ symmetry and has three inequivalent atoms. The 32-atom cluster has an unusual
cage-like structure which has an elongated structure. The symmetry for the As$_{32}$ cluster 
allows for cyclic permutations plus non cyclic permutations followed by inversion.
 The As$_{36}$ has D$_{6h}$ symmetry while the
60-atom cluster has a structure similar to the C$_{60}$ structure. However, the As$_{60}$ cluster is
much larger than the C$_{60}$ in size. Apart from the As$_{32}$, all the clusters from n=20 onward are
spherical in shape. While \ascage is made up of pentagons, the 28-, 32-, 36-, 60-atom clusters have
both pentagonal and hexagonal faces. From the Table \ref{table:str}, it becomes evident that
beyond As$_{20}$,  the bonds of the clusters are not of equal length and some the bonds become
stretched in the larger clusters. 

\begin{table}
\caption{ Optimized values of bond lengths, angles between inequivalent atoms and symmetry group 
for the As$_n$ cages }
\label{table:str}
\begin{tabular}{lccc}
\hline
Cage      & As-As (\AA)  & $\Theta$ (Degree)   &  Symmetry  \\
\hline

As$_4$    &  2.469       &   60.0             &    T$_d$     \\
As$_8$    &  2.547       &   90.0             &    O$_h$     \\
As$_{20}$ &  2.497       & 107.9 - 108.2$^o$  &    I$_h$     \\
As$_{28}$ & 2.462-2.524  & 105.4 - 120.9$^o$  &    T$_d$     \\
As$_{32}$ & 2.441-2.536  & 104.0 - 124.4$^o$  &    D$_3$     \\
As$_{36}$ & 2.444-2.538  & 107.2 - 130.4$^o$  &  D$_{6h}$ \\
As$_{60}$ & 2.427-2.737  & 108.0 - 120.0$^o$  &    I$_h$     \\
\hline
\end{tabular}
\end{table}

 The energetics of the clusters are shown in Table \ref{table:eng}. The atomization energy,
gap between the highest occupied and lowest unoccupied molecular orbitals (HOMO-LUMO), 
and the energies of dissociation into As$_4$ are presented in Table \ref{table:eng}. 
The binding energies per atom of the As clusters are centered around 2.7 eV. All the 
clusters considered here have close shell structures with a relatively large HOMO-LUMO gap 
and are Jahn-Teller stable. We have also confirmed that
all the clusters studied are local minima in the potential energy landscape.  The 
most striking aspect of these clusters is their dissociation energies with respect to As$_4$.  
Although the atomization energies and vibrational frequencies indicate the clusters to be stable, 
apart from the As$_{20}$ cluster, all the other cages are unstable with respect to As$_4$. It 
may be pointed out that Shen and Schaefer have predicted that As$_{20}$ to be competitive with 
As$_4$ clusters. On the other hand, the experimental mass spectrum of the
\nimolo clusters show peaks for all As$_x$Ni$_{12}$ clusters where x=1,21 indicating that in
this cluster the dissociation channel is As monomers rather than tetramers.  
This was shown due to the unusual change in bonding of the As$_{20}$ in the encapsulated form 
\cite{BZRP}. The stability of the \ascage cage compared to all the other cage structures 
is still surprising.

 The binding energy of the \ascage is slightly higher than the \as4 cluster. The
\as4 clusters are highly unreactive as evident from the large HOMO-LUMO gap. 
The atomization energy for \as4 is larger than other clusters 
while the HOMO-LUMO gap of \as4 is large which indicates the special structural and chemical
stability of the \as4. The \ascage is characterized  by even larger atomization energy which
indicates its stability with respect to \as4.
The dissociation energy for the tetramer channel for all the other clusters are large and positive.
The instability
of the clusters increases towards the larger sizes. One possible explanation 
can be the stretching of the bond-angle in the larger clusters. As can be seen from the Table
\ref{table:str}, the bond angles in the larger clusters of size n=28 onward are twice as 
large as those in the tetramer. 

  The dissociation of all the As cages except \ascage is exothermic with respect to \as4 units
and the energy released
lies between 1 to 11 eV. This is by no means large as in the case of nitrogen cages \cite{O} which
dissociates into N$_2$ units.
Although N and As occur in the same group, N$_2$ is more stable due to the formation of triple
bond than the As$_4$. It may be mentioned here that dissociation of nitrogen cages of size 4 to 
20 are found to 
release energy of the order of 3.0 to 3.8 Kcal/gm \cite{O}.

\begin{table}[b]
\caption{The atomization energies (AE), HOMO-LUMO Gaps and dissociation energies(DE) for 
the tetramer channel with
and without zero-point energy (ZPE)
are presented for As$_n$ where n= 4, 8, 20, 28, 32, 36, and 60. All 
values are in eV.}
\label{table:eng}
\begin{tabular}{lcccc}
\hline
Cage      & AE                   &  HOMO-LUMO         & DE      & DE including \\
          & (eV)                 &    gap (eV)        & (eV)    &  ZPE (eV) \\
\hline
As$_{4}$  & 2.76                    &   4.10             &  0.00   &  0.00 \\
As$_{8}$  & 2.62                    &   1.38             &  1.12   &  1.17 \\
As$_{20}$ & 2.79                    &   1.44             & -0.54   & -0.48 \\
As$_{28}$ & 2.70                    &   1.50             &  1.68   &  1.75 \\
As$_{32}$ & 2.69                    &   1.41             &  2.23   &  2.34 \\
As$_{36}$ & 2.67                    &   1.31             &  3.14   &  3.24 \\
As$_{60}$ & 2.57                    &   1.28             & 11.29   & 11.30 \\
\hline
\end{tabular}
\end{table}

\begin{figure}[t]
\epsfxsize=4.5in
\centerline{\epsfbox{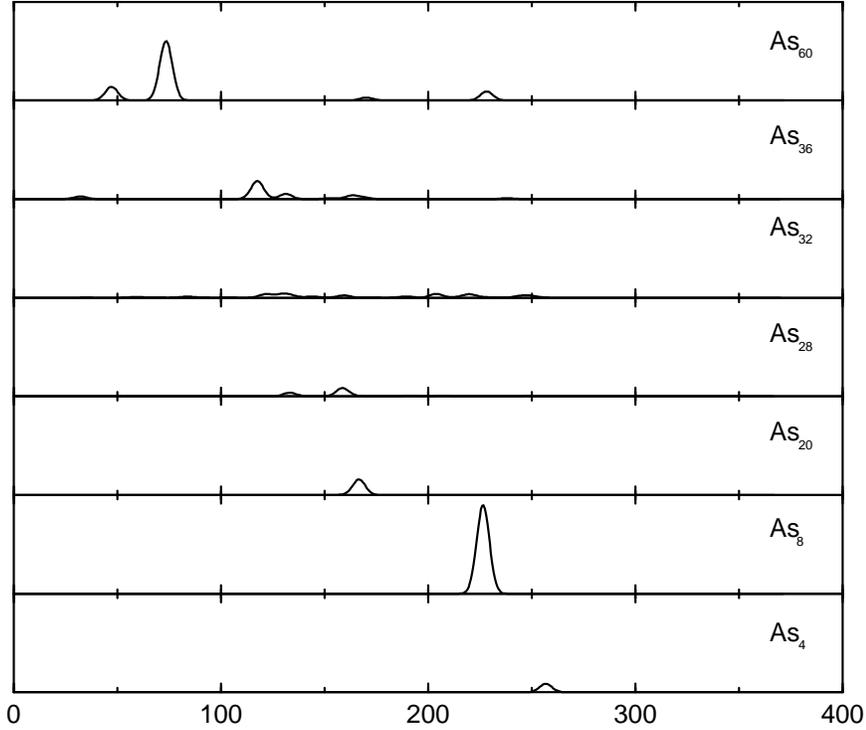}}
\caption{\label{IR}  IR spectra of   As$_{4}$,  As$_{8}$, As$_{20}$,  As$_{28}$, As$_{32}$ , As$_{36}$, and As$_{60}$ clusters.
}
\end{figure}

\begin{figure}[t]
\epsfxsize=4.5in
\centerline{\epsfbox{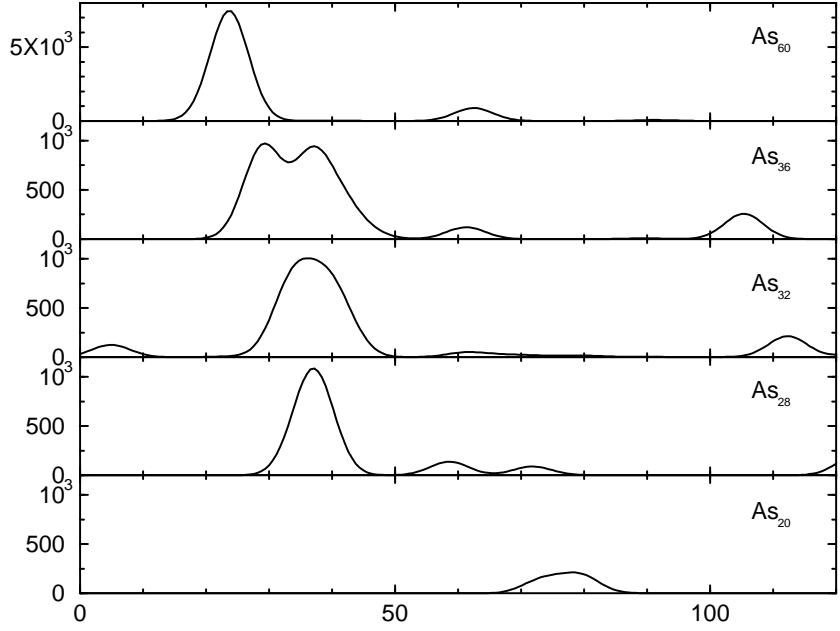}}
\caption{\label{Raman1}  Raman spectra of   As$_{20}$,  As$_{28}$, As$_{32}$ , 
As$_{36}$, and As$_{60}$ clusters in the frequency range 0-120 \cmin. The Raman spectra of the
As$_4$ and As$_8$ clusters do not have any peak in this range.  The Raman intensity of the As$_{60}$
is much larger compared to the other clusters and therefore is plotted in a different scale.
}
\end{figure}

\begin{figure}[t]
\epsfxsize=4.5in
\centerline{\epsfbox{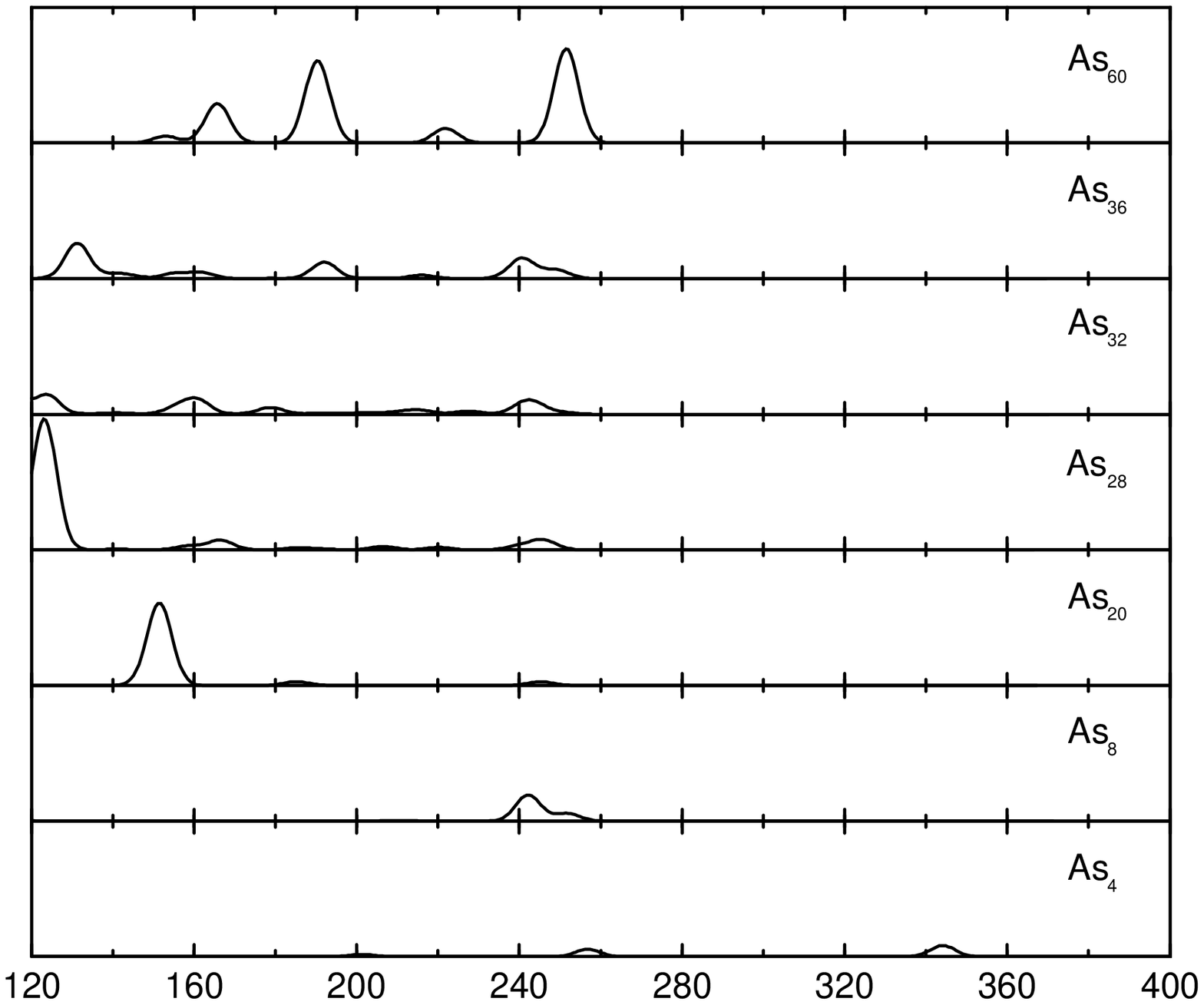}}
\caption{\label{Raman2}  Raman spectra of   As$_{4}$,  As$_{8}$, As$_{20}$,  As$_{28}$, As$_{32}$ , 
As$_{36}$, and As$_{60}$ clusters in the frequency range 120 - 400 \cmin.   All the panels show
Raman intensity in the same scale.
}
\end{figure}

The predicted infrared(IR) and Raman spectra are presented in Figs. \ref{IR}, \ref{Raman1}, and \ref{Raman2}. These spectra may help in experimental characterization
of such cages. The \as4, As$_8$ and \ascage show only one IR active mode each at 257 \cmin, 
227 and 167 \cmin. The larger cages have more dispersion in their
bond-lengths and hence in the spring constant resulting in more IR active modes.  One noticeable
feature is that for the more symmetric cages the IR active frequency goes down across the series.
Also while the peaks are high for As$_4$ and As$_8$, the intensities are much diminished for 
the higher size clusters. 

  The Raman spectra of the As cages are plotted in the frequency ranges of 0 - 120 \cmin and 
120 - 400 \cmin in Figs. \ref{Raman1} and \ref{Raman2} respectively. The larger clusters show 
high peaks in the low frequency region which are several
order of magnitude larger than that of the As$_4$ cluster. 
The Raman active modes are the ones which change the polarizability of the cages. 
The frequency of the highest Raman peak decreases as the number of atoms
and the radius of the cage increases. The polarizability of the As$_{60}$ is 43.4 per bohr$^3$  per
atom and its radius is 6.27 \AA. Due to the larger radii of the higher order clusters, the 
polarizability and its derivative increases which in turn influences the Raman intensity. This 
trend is seen across the series. 
The polarizabilities and the average radii of the clusters are shown in Table 
\ref{table:pol}. Since the polarizability is proportional to  the volume R$^3$,  the larger
clusters have higher polarizabilities. A comparison with C$_{60}$ fullerene shows \cite{quong} 
that in the
As$_{60}$ the polarizability is about 4-5 times larger. This is consistent with the radii of the 
two cages - the As$_{60}$ cage is about 2 times larger than the C$_{60}$ cage. The As$_{60}$ shows
an intense Raman peak at 24\cmin corresponding to a H$_g$ mode of vibration which is several times 
larger than those for As$_{28}$, As$_{32}$, and As$_{36}$. 
Unlike the IR spectra, the intensity of the highest peaks gets larger as the cluster size grows. 
For a plane polarized incident light and under the condition
that the direction of incident beam, the polarization direction of the incident light and the
direction of observation are perpendicular to each other, 
the Raman scattering cross-section  is given by the following equation \cite{C} : 
\begin{equation}
\frac{d \sigma_i}{d\Omega} =\frac{(2 \pi \nu_s)^4}{c^4} \frac{h(n_i+1)}{8 \pi^2 \nu_i} \frac{I_{Ram}}
{45}
\end{equation}
 where $\nu_s$ is the frequency of scattered light, $n_i$ is Bose-Einstein statistical factor, 
$\nu_i$ 
is frequency of the i$^{th}$ mode of vibration, and
\begin{equation}
I_{Ram} = 45 (\frac {d\alpha}{dQ})^2+ 7 (\frac{d\beta}{dQ})^2 = 45 \alpha''^2+ 7 \beta'^2 
\end{equation}
 where
\begin{eqnarray}
    \alpha''  & = & \frac{1}{3}(\alpha'_{xx}+ \alpha'_{yy}+\alpha'_{zz})  \nonumber \\
  \beta'^2   & = & \frac{1}{2}[(\alpha'_{xx}-\alpha'_{yy})^2 + 
             (\alpha'_{xx} -\alpha'_{zz})^2+     \nonumber    \\
             &   &(\alpha'_{yy} -\alpha'_{zz})^2+
             6(\alpha^{'2}_{xy}+ \alpha^{'2}_{xz}+\alpha^{'2}_{yz})] 
\end{eqnarray}
Here, $\alpha''$ and $\beta'$ are the mean polarizability tensor derivative and  the anisotropy of 
the polarizability tensor derivative respectively.  $I_{Ram}$ is the Raman scattering activity and 
Q is the normal mode coordinate. The scattering cross section is inversely proportional to the 
frequency of the
vibrational mode and therefore at low frequencies the Raman intensity increases as can be seen from
Fig. \ref{Raman1}.

The Raman scattering intensities in the frequency range 120- 400
\cmin are shown in Fig. \ref{Raman2}.  All the modes of As$_4$ show Raman activity since they are 
respectively of A$_1$, T$_2$ and E symmetry. The highest peak for As$_4$ occurs at 344 \cmin
in good agreement with experimental value of 356 \cmin \cite{RKH}. The experimental Raman spectra
of As vapor shows a strong peak at this frequency \cite{RKH}. The As$_4$ shows other peaks at
201 and 256 \cmin. The experimental spectrum displays a broad peak at around 250 \cmin which becomes
much broader at high temperature. A weighted Raman spectra of the most stable species, namely
As$_4$ and As$_{20}$ shows an asymmetric peak around 250 \cmin similar to the experimental profile.
Another peak at 200 \cmin is  less intense but is still visible in the experimental
spectrum. The calculated As$_{20}$ Raman spectrum shows prominent peaks at frequencies 
185, 151, and 79 \cmin. Since the experimental Raman spectrum is measured  between 120 to 450 \cmin,
the large signatures of the \ascage can not be discerned from the experimental spectrum. 
In the high density limit, the peaks of the weighted spectrum
shows small peaks at 150 and 79 \cmin whereas around 250 a somewhat broad peak is 
observed. In the
low density limit, the peaks below 150 becomes much more prominent than the peaks at higher 
frequencies. An experimental measurement of the Raman spectra in the low frequency region can be
helpful in identifying the existence of As$_{20}$.

\begin{table}[t]
\caption{The average radii (\AA) and average polarizabilities in  \AA$^3$ of the As cages. }
\label{table:pol}
\begin{tabular}{lcc}
\hline
Cage      &  Radius (\AA)   &  Polarizability (\AA$^3$)     \\
\hline
As$_{4}$  &   1.51   &   18.05                  \\
As$_{8}$  &   2.21   &   35.69                  \\
As$_{20}$ &   3.50   &   96.01                  \\
As$_{28}$ &   4.20   &  145.19                  \\
As$_{32}$ &   4.49   &  170.18                  \\
As$_{36}$ &   4.76   &  196.67                  \\
As$_{60}$ &   6.29   &  386.41                  \\
\hline
\end{tabular}
\end{table}

In conclusion, 
we
have studied the geometry, vibrational stability, energetics and IR and Raman spectra of As cages of size
n = 4, 8, 20, 28, 32, 36, and 60 to examine the possibility of existence of As cages. 
We find that all the clusters except \ascage  are unstable against dissociation into As$_4$ units. 
The energy released in the 
exothermic dissociation is significantly smaller then that associated with the isoelectronic 
nitrogen clusters.  We determine the  vibrational stability of the clusters and also 
predict the IR and Raman spectra.  The polarizability increases with cluster size. 
We expect our study will
inspire experimental search for such metastable clusters and suggest that the Raman peak at 
roughly 80 \cmin could be used to identify the existence of As$_{20}$.

  TB and MRP acknowledge financial support from ONR (Grant No. N000140211046) and
by the DoD High Performance Computing CHSSI Program. RRZ thanks GMU for support and MRB 
thanks CONACYT 40393-F for support.

\end{document}